\documentclass[twocolumn,aps,prl,showpacs]{revtex4-1}
\usepackage{graphicx,epsf,epsfig}
\usepackage{dcolumn}
\usepackage{bm}
\usepackage{amsmath, amsthm, amssymb}
\usepackage{color}
\usepackage{ulem}
\usepackage{epstopdf}
\definecolor{pink}{rgb}{1,0.5,0.5}

\begin{document}

\title{Low-bias mobility vs high-bias saturation current in graphene based transistors}
\author{F. Tseng and A.W. Ghosh}
\affiliation{Dept. of Electrical and Computer Engineering, University of Virginia, Charlottesville, VA 22904}%

\begin{abstract}
We describe the fundamental trade-offs in engineering the mobility, current
saturation and ON-OFF ratios in graphene transistors. Surprisingly, the trade-offs
arise solely from an {\it{asymptotic constraint}} on the high energy bandstructure and \textit{independent of scattering processes}. This places graphite
derivatives (bulk monolayer graphene, uniaxially strained graphene nanoribbons, carbon nanotubes and bilayer graphene) on the same 3-parameter mobility-bandgap-scattering length ($\mu-E_{gap}-\lambda$) plot, proximal to other semiconductors.  In addition to this low-bias trade-off, the high bias current bears signatures of the underlying saturation mechanism, arising through phonon scattering or $\Gamma$-point  suppressed density of states opening  bandgap.\end{abstract}


\maketitle

The incredible properties of graphene have stimulated intense exploration into its potential as
an electronic device \cite{Novo}. Graphene retains many admirable properties of nanotubes, such as
a high mean-free path for scattering and a paucity of surface dangling bonds, minus many of their
disadvantages, such as insensitivity to atomic chirality and the possibility of top-down lithographic
patterning. This has led to the study of graphene devices that range from evolutionary (such as
monolithically patterned wide-narrow-wide ribbons \cite{wnw}) to revolutionary (such as Veselago
`waveguides' and switches\cite{Veselago} , and Bilayer pseudospin field effect transistors or BisFETs \cite{swan}).

The biggest problem for a graphene-based electronic switch is undoubtedly its narrow band-gap. Various efforts to
engineer the band-gap have consistently seen gaps narrower than $\sim 200$ meV.
Such a small band-gap threatens to severely limit the room temperature ON-OFF ratio to only $\sim$ 50 (keeping in mind that for switching
applications the drain and gate bias voltages are typically comparable). The small bandgap also generates a large subthreshold swing $\sim$
145 mV/decade \cite{iannacone} and large source-to-drain tunneling currents at modest voltages \cite{tseng_apl}. 

The reliance on chemical means or quantization for band-gap creation compromises overall device scalability.
More seriously, the bandstructure of graphitic derivatives pose a stringent restriction on their electronic properties,
so that opening a band-gap via strain or chemistry (as in graphane \cite{graphane}) undermines the incredible
mobility of graphite electrons arising from their small effective mass.

\begin{figure}[t!]
\centering
{\epsfxsize=2.6in\epsfbox{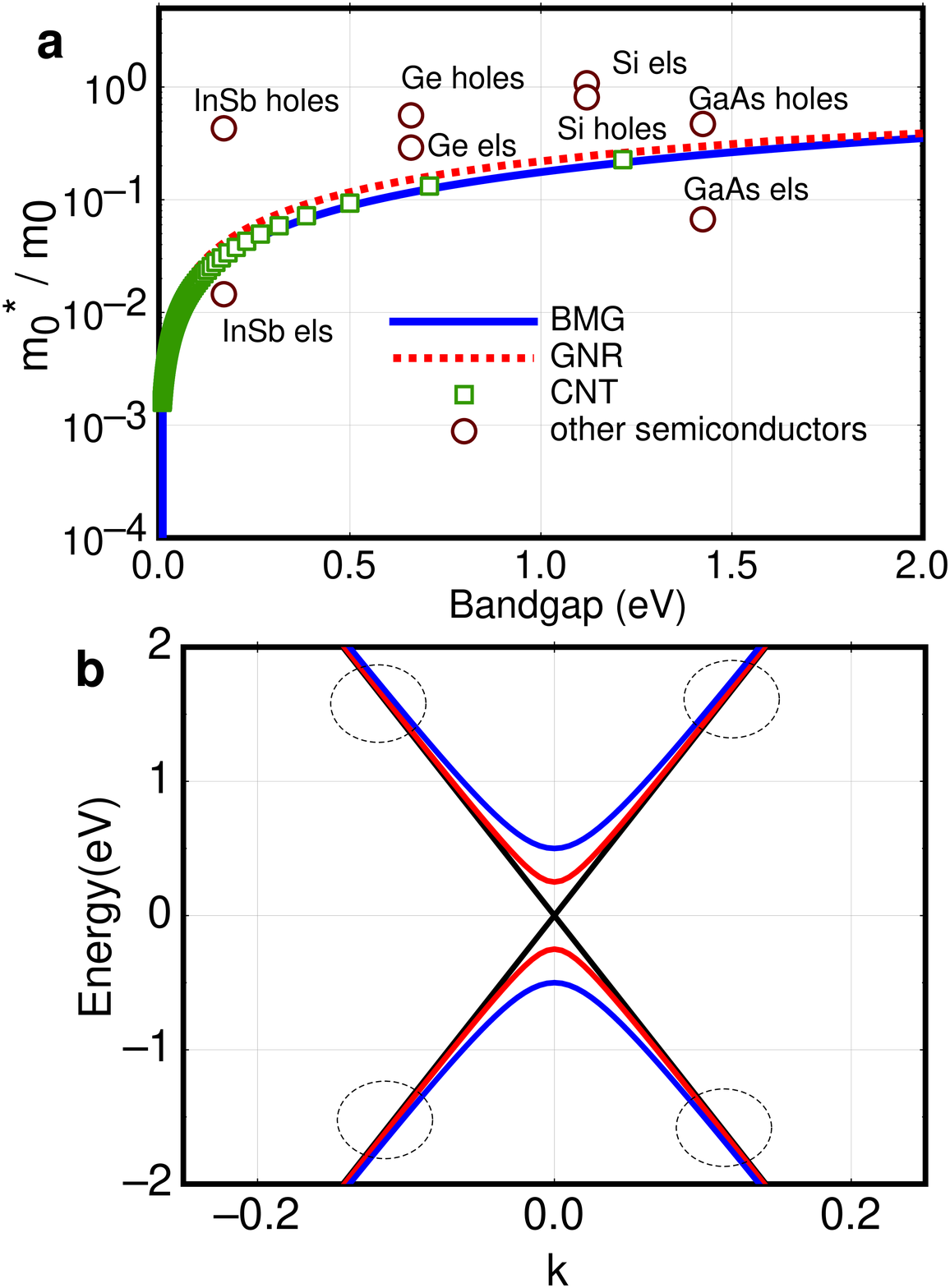}}
\caption[]{ Regardless of the bandgap opening mechanism, all monolayer graphitic materials such as SLG, CNTs, GNRs are constrained to a single (a) effective mass versus bandgap curve. Graphitic materials are fundamentally limited by a (b) dispersion relation that is pinned to its linear form at high energies. Compared to other common group IV and III-V semiconductors at corresponding energy bandgaps, graphene electron carriers are not much lighter.}
\vspace{-.35cm}
\end{figure}

In this paper we explore fundamental {\it{band-related constraints}} on both the low-bias mobility and
the high-bias current saturation in FETs created out of graphite-based materials.
We outline an intimate trade-off between mobility (switching speed) and ON-OFF ratio (bit error
rate) that stems from an {\it{asymptotic constraint}} on the high-energy bandstructure
of all the well known graphite derivatives, including single layer graphene (SLG), strained graphene nanoribbon (sGNR), carbon nanotubes (CNT) and bilayer graphene (BLG), \textit{even without a corresponding change in scattering length}.
Lastly, we outline different mechanisms behind current saturation in graphene, prior to band-to-band
tunneling -- focusing on discernible inelastic electron tunneling spectroscopy(IETS)~\cite{iets1} signatures arising from saturation due to the
$\Gamma$-point, bandgaps and phonon scattering.

\textit{Relating mobility to bandgap.} Irrespective of the underlying bandgap opening mechanism, the bandstructure of all graphite derivatives reach a linear dispersion at higher energies as expressed in  $E(k)\approx\pm\sqrt{(E_{gap}/2)^{2}+ (\hbar v_0 k)^{2}}$, where $\hbar$ is reduced Planck's constant. For SLG, \textit{v}$_0 = {3a_0t}/{2\hbar}$ is the velocity of the high energy graphene electrons $\sim 10^8$ cm/s, with $a_0$ being the C-C bond-length and $t \sim 2.5$ eV being the bond energy. Whether this slope itself changes depends on the wavelength of the perturbing potential (e.g. confinement or strain) relative to the graphite lattice constant.
More specifically,
\begin{equation}
\label{dispersion}
E = \begin{cases}
\pm\displaystyle\sqrt{ (E_{c})^{2}+ (\hbar v_{o} k)^{2}}  , & \mbox{(s)GNR} \\
\pm\displaystyle\frac{3a_0t}{2}\sqrt{\displaystyle\Biggl(\frac{2}{3d}\Biggr)^2+k_x^2},& \mbox{CNT} \\
\pm\displaystyle\sqrt{A + \hbar^2v_0^2k^2+(-1)^\alpha\sqrt{B + C\hbar^2v_0^2k^2}},  & \mbox{BLG}\\
\end{cases}
\end{equation}

\parindent0ex where $E_{c}$ and $\hbar v_{o}$ from sGNR (1) are defined as
 
\begin{equation}
\label{sG_E}
E_{c} = \begin{cases}A'(\gamma_{1}+\gamma_{3})+2B'\{\gamma_{1}cos\frac{p\pi}{N+1} \nonumber\\ + \gamma_{3}\left[C'+\left(1-C'\right)cos\frac{2p\pi}{N+1}\right]\}
\end{cases}
\end{equation}
\vspace{-1cm}

\begin{equation}
\label{sG_hv}
\hbar v_{o} = \begin{cases}
(3d)^{2}\{ -\frac{1}{2}s\gamma_{1}B'cos\frac{p\pi}{N+1}[A'(\gamma_{1}+\gamma_{3})\nonumber\\+2B'\gamma_{3}\left(C'+\left(1-C'\right)cos\frac{2p\pi}{N+1}\right)] \nonumber\\-\gamma_{3}[A'\gamma_{1}+(A'-1)\gamma_{3}\nonumber\\+2B'\gamma_{3}\left(C'+\left(1-C'\right)cos\frac{2p\pi}{N+1}\right)]\} 
\end{cases}
\end{equation}

\parindent0ex Meanwhile the terms for sGNR~\cite{strain} dispersion, A', B', C', $\gamma_1$, $\gamma_3$, p, and (N+1) are defined in the reference~\cite{sGterms}. Setting appropriate strain terms to zero recovers the dispersion for GNRs with first and third nearest neighbor interactions, and edge distortion\cite{white}. For BLG  the terms A, B, C, and $\alpha$ are defined in reference~\cite{bilayer_def}. 
{\it{Since the opening of $E_{gap}$ does not influence its high energy sector that is pinned to the linear dispersion, it flattens the band-curvature and increases its effective mass}} (Fig. 1). There is thus an inherent trade-off between ON-OFF ratio (bandgap) and mobility (effective mass) {\it{based on bandstructure considerations alone}}.  Taking the second derivative at k = 0 gives the effective mass at the band-bottom, $m^{*}_0= {E_{gap}}/{v_0^{2}}$, implying that the kinetic energy gained by the electrons and holes equals the energy lost from the crystal potential during band-gap opening. The definition of effective mass away from the band-bottom requires careful consideration. One can write the mobility $\mu={q\lambda}/ {m^{*}v}$, where the effective mass $m^{*}$ and the carrier velocity {\it{v}} are in general, energy-dependent, and $\lambda$ is a constant mean free path.  Invoking the dynamical definition of effective mass, $\mu$ can be rewritten in terms of the gate voltage by first replacing {\it{$m^{*}$v}} with $\hbar{\it{k}}_F$ at the Fermi energy (i.e., using $m^* = p/v$ reduces to $m^*_0$ only at the band-bottom~\cite{m_eff}). From the graphene dispersion, ${\it{k}}_F={1}/{\hbar v_0 }$$\sqrt{{E_F^{2}}-{{E_{gap}^{2}} /{4}}}$, where ${\it{E_F}}={E_{gap}}/{2}$+$qV_{G}$ for the electronic sector. We then have
\begin{equation}
\mu=\frac{q \lambda v_0 }{\sqrt{qV_{G}E_{gap}+(qV_{G})^{2}}}
\end{equation}
Since $V_G$ also determines the 2-D electron density through $k_F = \sqrt{\pi n_{2D}}$, the gate dependence
of the mobility translates to a dependence on $n_{2D} =
V_G(V_G + E_{gap})/L\pi\hbar^2v_0^2$ for a channel of length $L$.

\begin{figure}[b]
\vspace{-0.25 cm}
\centering
{\epsfxsize=2.6in\epsfbox{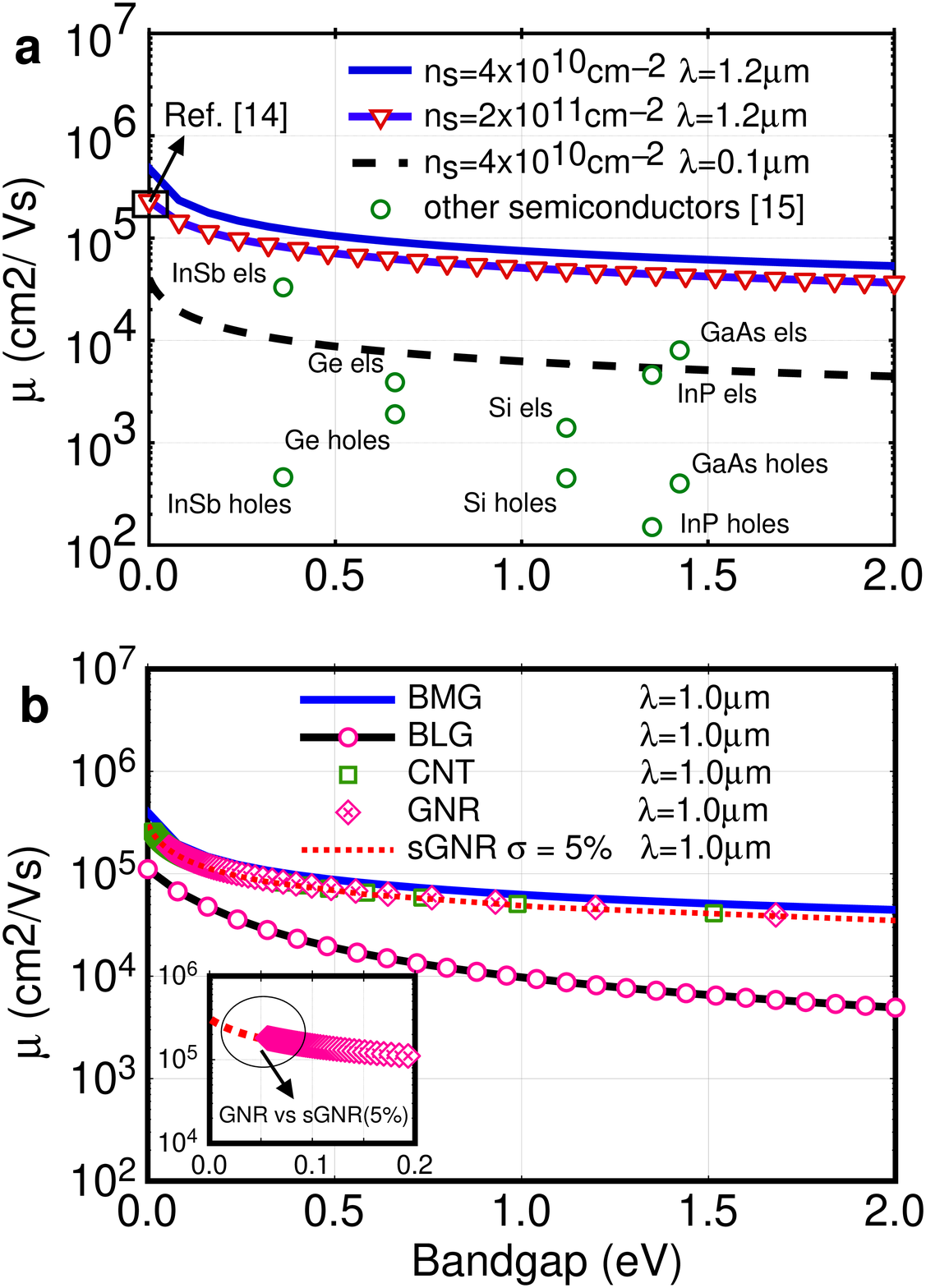}}
\caption[]{ 3- parameter ($\mu-E_{gap}-\lambda$) illustrates the fundamental cost in opening a bandgap to carrier mobility in graphitic materials. With an electron density of $2$x$10^{11} cm^{-2}$ and $\lambda\approx 1.2\mu m$, zero-bandgap graphene reaches a mobility of $230,000 cm^{2}/Vs$ matching experiments~\cite{gMobile}. Compromising $\lambda$ as mentioned in the text potentially lowers carrier mobility comparable to some group IV and III-V semiconductors~\cite{Sze} for a given $E_{gap}$. (b) All monolayer graphitic derivatives fall in a single curve with small difference in the value of $v_0$~\cite{white}. For the same set of armchair ($3p+1,0$) GNRs, we show in the subplot that uniaxial strain($\sigma=5\%$) can close bandgaps(extend $\mu$). BLG trend deviates from monolayer graphitic derivatives due to its non-linear zero-bandgap dispersion and its unique band-bottom features in the presence of a bandgap~\cite{bilayer, Ohta}.  }
\vspace{-0.45 cm}
\end{figure}

\parindent3ex While much interest in graphene, such as for RF applications, stems from its ultra-high mobility~\cite{gMobile}, for digital switching applications one gets a more complete picture by plotting an $E_{gap}$ vs $\mu$ curve for various $\lambda$s, as in Fig.2. One can see that for zero band-gap at room temperature, we get the maximum theoretical mobility $q\lambda v_0/kT \sim 400,000$ cm$^2/Vs$ for a $1.2 \mu$m scattering length.
The perceived advantages of graphene get compromised when one factors in both its switching speed and its ON-OFF ratio.  Scattering processes from impurities \cite{Sarma, JenaMob}, bulk or interfacial phonons \cite{Goldsman, JenaMob}, and line edge roughness~\cite{RmTemp,Yang,JenaMob} in GNRs  are expected to shorten $\lambda$, further decreasing the mobility in addition to the fundamental bandstructure constraints. From our analysis it seems that once the advantages of low $m^*$ and high $\lambda$ are compromised, graphitic switches act comparable if not worse than Indium Antimonide (InSb) and Gallium Arsenide (GaAs), and marginally better than Silicon (Si). Using the dispersions (1) and the generic ($\mu-E_{gap}-\lambda$) formulation (2), we show universality in trade-offs amongst single layer graphitic derivatives (Fig.2b). 
We would emphasize the trade-offs in opening $E_{gap}$ on $\mu$ is synonymous with conductance($G=\frac{2q^{2}}{h}M(E)T\frac{\lambda}{\lambda+L}$), where the \textit{mode density (M(E)) near band-bottom is diluted independent of} $\lambda$, while transmission per mode(T) and graphene length (L) are constants.

\textit{High bias current: Modeling scattering.} We saw that the low-bias mobilities of all graphitic derivatives are intimately connected with their band-gaps. A similar universality arises for the high-bias current. The diminishing density of states for graphene near the $\Gamma$ point conspires together with various scattering mechanisms to create a current-voltage (I-V) characteristic that shows a tendency to saturate, followed sharply by a rise due to band-to-band tunneling. A closer look at the I-V and its derivatives provides useful insights into the underlying scattering mechanism. To this end, we will now discuss how saturation occurs in our model for graphene FETs.

Electron transport through a nanoscale object is described using the non-equilibrium Green's function (NEGF) formalism that simplifies to Landauer theory for coherent quantum flow \cite{Datta},
\begin{eqnarray}
I = \frac{2q}{h}\int dET(E)(f_1-f_2)
\end{eqnarray}
where $f_{1,2}$ are the bias-separated contact Fermi-Dirac distributions, the transmission $T = trace(\Gamma_1G\Gamma_2G^\dagger)$, with the Green's function $G = (ES - H -\Sigma)^{-1}$, $S$, $H$ and $\Sigma$ being the channel overlap matrix, Hamiltonian and self-energy matrices respectively, 
and $\Gamma_i = i(\Sigma_i-\Sigma_i^\dagger)$ is the level broadening. A recursive technique allows us to get
$\Sigma$ for any layered contact structure in the non-interacting limit. The influence of scattering modifies the Green's function through Dyson's equation:
\begin{equation}
G^{-1}=G_{0}^{-1}-\Sigma_S,
\end{equation}
where $G_{0}$ is the unperturbed response and $\Sigma_S$ represents perturbing interactions from scattering processes such as from impurities (elastic) or phonons (inelastic). From an energy dependent density of states $D(E)=\frac{|E|S}{2\pi\hbar^{2}v_{0}^{2}}$$\left[\Theta(E-E_{c})+\Theta(E_{v}-E)\right]$ we can extract the imaginary part of $G_{0}$ and Hilbert transform to get the real part as well. $S$ is the area of the SLG sheet (500nm x 500nm).
For inelastic phonon scattering, we can invoke the self-consistent Born approximation as follows:\begin{eqnarray}
\Sigma^{in,out}(E) &=& \Sigma^{in,out}_{1}(E)+ \Sigma^{in,out}_{2}(E)+\Sigma^{in,out}_{ph}(E) \nonumber\\
G^{n,p}(E) &=& G(E)\Sigma^{in,out}(E)G^\dagger(E) \nonumber\\
\Sigma^{in,out}_{ph}(E) &=& {\cal{D}}_0(\omega)\otimes\Biggl[N_\alpha(\omega)G^{n,p}(E\mp\hbar\omega) \Biggr.\nonumber\\
\Biggl. &+& \left(N_\alpha(\omega)+1\right)G^{n,p}(E\pm\hbar\omega)\Biggr] \nonumber\\
I_1 &=& \frac{q}{h}\int dETr[\Sigma^{in}_1G^p - \Sigma^{out}_1G^n],
\end{eqnarray}
where $\Sigma^{in,out}_{1,2}(E)$ are the self-energy terms for the contacts and $\Sigma^{in,out}_{ph(E)}$ is the self-energy term for the inscattering and outscattering that occur at a specific energy (E=$\pm\hbar\omega$). The scattering deformation potential, ${\cal{D}}_0(\omega)$=$\lambda_{epc}\hbar\omega$~\cite{phonon_run}, where $\lambda_{epc}$ is the electron-phonon coupling constant (epc). For acoustic phonons we use a Debye model to cover the range of frequencies up to the Debey frequency. $N_\alpha(\omega) = [e^{\hbar\omega_\alpha/k_BT}-1]^{-1}$ is the phonon Bose-Einstein distribution describing the population of phonons at $\hbar\omega$, and ${\it{I_1}}$ is the terminal current.

\begin{figure}[b!]
\centering

\vspace{-.45 cm}
\hspace{-1.6cm}
{\epsfxsize=4.0in\epsfbox{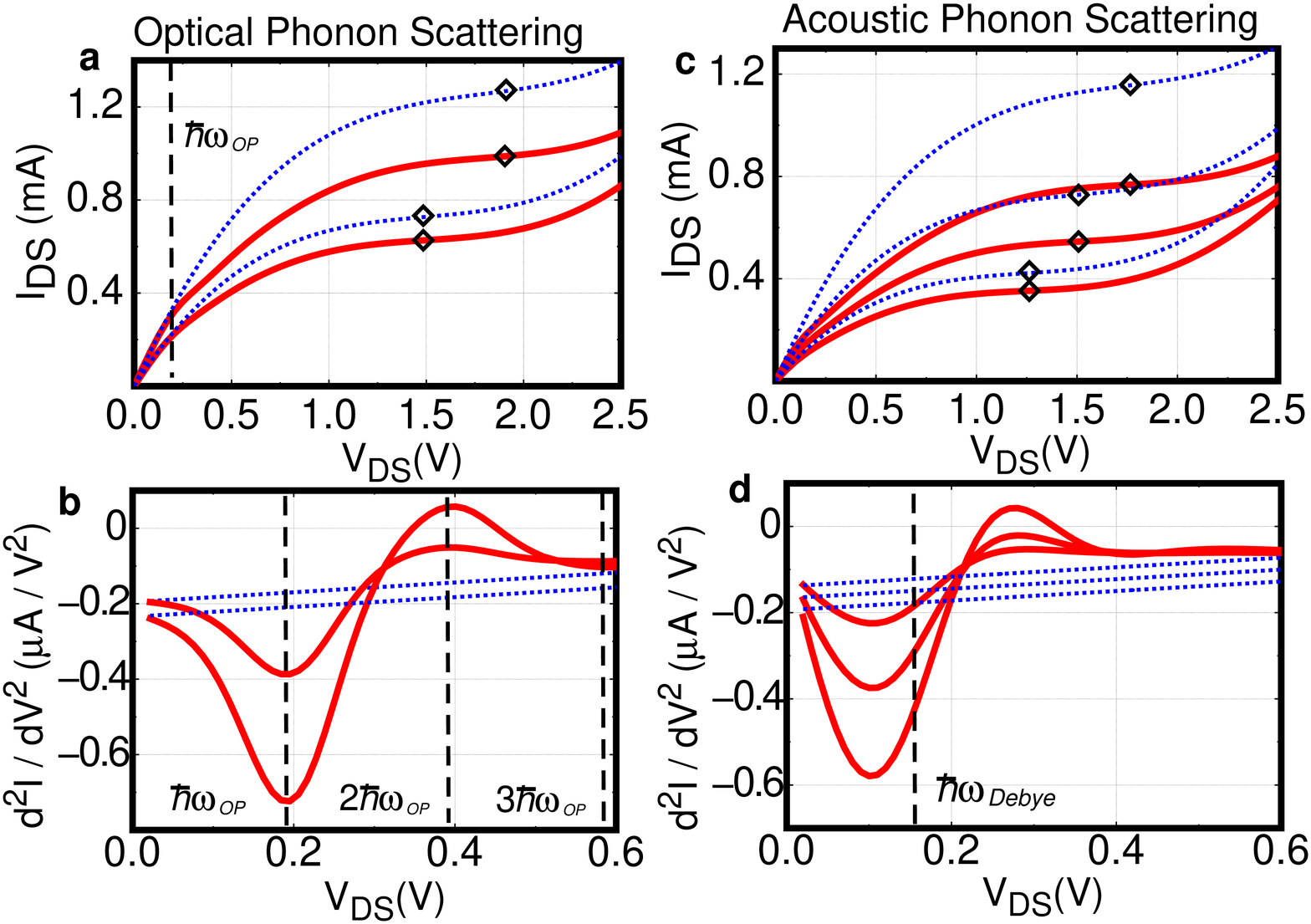} }
\caption[]{IV-characteristics and IETS for scattering induced saturation (SIS) in zero-bandgap gFETs (red-curves). Ballistic case in dashed-blue curves.  (a) LO phonons in the IV ($V_G=1.5V, 1.8V$) are active at $\hbar\omega\approx0.19eV$. (b) Vibronic signatures in IETS at 0.19eV are seen with overtones at higher harmonics. (c) LA phonons are active over a range of energies from $\hbar\omega\approx0eV-0.16eV$ and contribute to  decrease in conductance ($V_G=1.25V, 1.50V, 1.75V$) (d) Corresponding IETS shows a broadened valley near the Debye frequency for LA modes.   *Parameters : $\Gamma_i\approx 2eV$~\cite{gamma}, graphene area (S): 500nm x 500nm, LO phonon $\lambda_{epc} \approx 0.804$, LA phonon $\lambda_{epc} \approx 0.40$ . Note different voltage ranges in IETS (b,d).}

\vspace{-0.35cm}
\end{figure}

\textit{Current Saturation Mechanisms.} Prior to band-to-band tunneling in a gFET is an onset of scattering-induced-saturation (SIS) due to phonons from lattice vibrations either inherent to graphene or the underlying substrate atoms. We focus on intra-valley longitudinal optical (LO) and longitudinal acoustic (LA) phonons as examples that strongly couple with electrons and influence transport. Using the Einstein model LO phonons, are active at single energy $\hbar\omega\approx0.19eV$, while LA phonon are approximated with the Debye model are active across a bandwidth of energies ($\hbar\omega\approx0eV-0.16eV$).

Optical phonons are known to saturate the current in metallic CNTs depending upon on the strength of ${\cal{D}}_0(\omega)$ \cite{YaoSat}. ${\cal{D}}_0(\omega)$ for both LA and LO phonons from CNTs~\cite{Sayed} are applied to our scattering model for SLG without loss of generality to the onset of SIS on current. Accordingly, the onset deviation from the ballistic-IV is perceived at energies where phonons are activated (Fig.3 ).

\begin{figure}[t!]
\centering

\vspace{-.15 cm}
\hspace{-0.6cm}
{\epsfxsize=3.6in\epsfbox{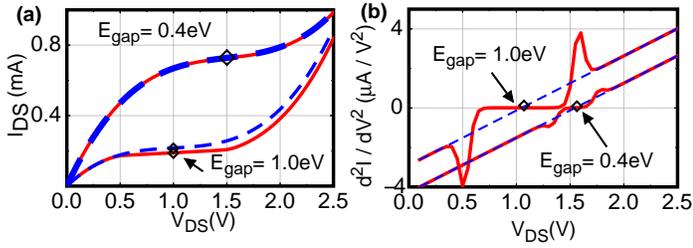} }
\vspace{-0.40cm}
\caption[]{ Narrow graphene bandgaps difficult to discern in the IV create a well defined plateau in the IETS. Zero-bandgap gFET in absence of SIS has a linear IETS from the symmetric inflection in the IV at the $\Gamma$-point. The diamond symbols mark the $\Gamma$-point shift at different levels of electrostatic doping($E_{gap}=0.4eV, V_G=1.5V; E_{gap}=1.0V, V_G=1.0V $ ).}

\vspace{-0.45cm}
\end{figure}

{\it{We thus find that a critical feature in LO phonon induced saturation is that the onset voltage for saturation is pinned to a gate-independent value given by the phonon frequency}}.
This feature can be accentuated by measuring the second harmonic response and plotting the curvature of the I-V curve, as in IETS measurements. A sharp valley in the $d^2I/dV^2$ curve is expected to arise at a gate independent voltage for LO phonon SIS (Fig. 3b). In addition, higher order harmonic side-bands in the IETS (Fig. 3b) are a visible and reminiscent feature in CNTs~\cite{phonon_run}. Meanwhile, the strength of LA phonons increases with $D(E)$ and are averaged over a range of frequencies creating a broadened valley in the IETS whose minimum is short of the Debye frequency (Fig. 3d). 

However, we add that opposite effects of scattering is visible in weakly coupled contact-channel gFET. Low bias regimes with a paucity of states near the $\Gamma$-point are prone to forward scattering which enhances conduction and further evidenced by peaks in the IETS at the relevant vibronic frequency. The opposite effect prevails at higher bias regimes with more states.

For zero-bandgap graphene, the paucity of states near the $\Gamma$-point creates a seemingly short plateau that is really an inflection in the IV. We denote the linearly decreasing $D(E)$ near the $\Gamma$-point at low bias regimes as the source of $\Gamma$-point induced saturation (GIS). To achieve currents independent of voltage requires a bandgap in graphene. However bandgap induced saturation (BIS) for even narrow bandgap graphene is nearly indiscernible from the IV alone due to band-to-band tunneling. Although difficult to measure, IETS should still resolve a short plateau corresponding to a narrow bandgap (Fig.4). 

\textit{Conclusion.}  In this work we point out that attempts to engineer a bandgap in gFETs invariably reduce their mobilities, based on asymptotic band-related constraints alone. On one hand, gFET as a switch is fundamentally compromised by $\mu-E_{gap}-\lambda$ relation. On the other hand, subthreshold conduction from band-to-band tunneling in narrow -bandgap (higher-mobility) gFETs compromises ON-OFF current ratios, thus requiring a careful balance in future design considerations. Finally, we differentiate IV saturation in gFETs arising from bandgaps, $\Gamma$-point and vibronic scattering. Based on these studies, we believe it should be straightforward to create a compact model for IVs based on any gFETs, that can incorporate a multitude of geometrical and atomistic variations as well as underlying physical scattering mechanisms.

\section{Acknowledgements}
We thank useful discussions with Supriyo Datta, Mikiyas Tsegaye, Mircea R. Stan, Kieth Williams, Kurt Gaskill and Jeong-Sun Moon. This work is supported by a NSF-NIRT and UVA-FEST awards.

\bibliography{graph_tradeoff_revtex.bbl}

\begin{thebibliography}{10}
\bibitem{gamma}
$\Gamma_i$=2eV comes from $\Gamma_i$ =$\hbar$ $v_{sat}/a$ where a= 0.142nm and
  $v_{sat}$ is the velocity saturation from the reference~\cite{Shepard}.
\end{thebibliography}
\end{document}